\documentclass[a4paper, amsfonts, amssymb, amsmath, reprint, showkeys, nofootinbib, twoside]{revtex4-1}
\usepackage[english]{babel}
\usepackage{graphicx}
\usepackage{subcaption}
\usepackage[utf8]{inputenc}
\usepackage{booktabs}
\usepackage[table,xcdraw]{xcolor} 
\usepackage{booktabs} 
\usepackage[colorinlistoftodos, color=green!40, prependcaption]{todonotes}
\usepackage{amsthm}
\usepackage{mathtools}
\usepackage{physics}
\usepackage{xcolor}
\usepackage{graphicx}
\usepackage[left=23mm,right=13mm,top=35mm,columnsep=15pt]{geometry} 
\usepackage{adjustbox}
\usepackage{placeins}
\usepackage[T1]{fontenc}
\usepackage{lipsum}
\usepackage{csquotes}
\usepackage[pdftex, pdftitle={Article}, pdfauthor={Author}]{hyperref} 

\begin{document}
\title{State transfer analysis for linear spin chains with non-uniform on-site energies}

\author{C.C. Nelmes, I. D'Amico, and T.P. Spiller}\email{timothy.spiller@york.ac.uk}
    \affiliation{Department of Physics, University of York, York, YO10 5DD, United Kingdom}
\date{\today} 

\begin{abstract}
High fidelity state transfer is an important ingredient of distributed quantum information processing. We present and analyse results on perfect and quasi-perfect state transfer with linear spin chains incorporating non-uniform on-site energies. The motivation is maintenance of coupling uniformity, which could be beneficial for some physical implementations. We relate this coupling uniformity to a particle in a discrete potential analogue. Our analysis further considers the statistical variation in couplings and on-site energies, as a function of increasing chain site number $N$. 
\end{abstract}


\maketitle
\section{Introduction} \label{sec:Intro}
The pursuit of efficient and reliable methods for transmitting information within quantum systems, whilst mitigating disruption in-transit, persists as a significant challenge within the field of quantum information. Spin chains, in the form of one-dimensional arrays of coupled spin-$\frac{1}{2}$ systems or qubits, have been proposed as ideal candidates for channels or `quantum wires'\cite{Bose_2003,Bose_2007, Nikolopoulos2}. For spin chains, there has been extensive work on the pre-engineering of interactions and reliance on the natural dynamics within spin networks, to accomplish perfect state transfer (PST) of an initial encoding of information \cite{Christandl1, Albanese, Christandl2, AllCoup, K-S, Nikolopoulos1,Yung,Vinet, Kay}. Prior research in this area has focused largely on control of the inter-qubit coupling interactions along the chain, whilst generally neglecting use of on-site/local magnetic field contributions, as a means to construct sufficient conditions for PST. Within this pursuit, models with prescribed coupling schemes have been proposed and can be shown to demonstrate periodic dynamics with fidelities close or equal to unity \cite{Christandl1, Albanese, Christandl2, AllCoup, K-S, Nikolopoulos1}.

 As will be discussed, utilisation of the natural system dynamics to accomplish PST, or quasi-PST (QPST) \cite{Vinet2, Godsil, Severini, Van Bommel}, in effect requires specification of the system Hamiltonian to deliver the desirable evolution. 
In this work we discuss examples from two different approaches to constructing suitable Hamiltonians, to achieve the desirable goals of both PST/QPST and coupling uniformity. Using a genetic algorithm (GA) approach \cite{ES,Luke,BN, LC} has the advantage of enabling exact coupling uniformity to be fixed, but then requires optimisation of QPST towards PST with a suitable fitness function for the GA \cite{Luke}. Alternatively, since exact spectral conditions for PST are known, use of inverse eigenvalue methods \cite{Yung, Kay, Vinet, DeBoor, Hochstadt, Hald,Parlett, Gladwell} can guarantee PST, but then require choices of spectra that give good (but cannot guarantee perfect) coupling uniformity. In both approaches, non-uniformity in the on-site energies is enabled, as for some physical implementations this may be preferable to non-uniformity in the couplings.

Some of the previous work on PST in spin chains, generally with uniform  on-site energies, was described via a group theoretical approach employing Lie algebra (SU(2)) of quantum angular momentum theory \cite{Christandl1,Albanese}. Within this framework, the coupling (or hopping) off-diagonal terms of the Hamiltonian tridiagonal matrix describing the linear chain, are related to Clebsch-Gordon coefficients resulting from the angular momentum raising ($L^+$) and lowering ($L^-$)operations. This conceptual mapping provides some physical intuition for the dynamics, through correspondence between a single spin-flip moving through a defined ($2L+1$)-site spin chain and a fictitious large spin (with $L = \frac{1}{2}(N-1)$) undergoing periodic motion in a transverse magnetic field \cite{Christandl1, Albanese, Christandl2, K-S}.
This angular momentum picture does not provide the same physical intuition for scenarios where the chain couplings are uniform, but the on-site energies are non-uniform. Therefore,  here, we will present a different conceptual framework, in which the PST/QPST spin-flip dynamics in the chain are related to the motion of a quantum particle in a finite discrete lattice, subject to a potential.

This paper is constructed as follows. The next section introduces the spin chain Hamiltonian and relevant properties for PST/QPST dynamics. This is followed by specific examples generated from a GA approach, where the couplings are fixed to be uniform. We then introduce the particle in a finite discrete lattice framework, along with the physical intuition that this provides. We follow this with other PST examples, generated from inverse eigenvalue methods. As these results do not have coupling uniformity imposed, we present detailed analysis of the variations in coupling and on-site energies, as a function of increasing chain length. Finally, we present further discussion on experimental implementations and then conclude.

\section{The spin chain model}\label{sec:Hamiltonian}
There is a formal equivalence between a 1/2-spin system and a qubit. Using the encoding
\begin{equation*}
|\downarrow\hspace{0.1cm}\rangle \rightarrow |0\rangle\\
\quad \& \quad
|\uparrow\hspace{0.1cm}\rangle \rightarrow |1\rangle,
\end{equation*}
the linear spin chain systems that we consider are described by the so-called XY model
\begin{align}
\Hat{H}_{XY} = & \sum_{i=1}^{N-1} J_{i,i+1}\left(|1\rangle \langle0|_{i} \otimes |0\rangle \langle1|_{i+1} \right. \nonumber \\
         & \left. + |0\rangle \langle1|_{i} \otimes |1\rangle \langle0|_{i+1}\right) + \sum_{i=1}^{N} \varepsilon_{i} |1\rangle \langle1|_{i}.
         \label{Hami}
\end{align}
Here, $N$ is the number of sites,  $J_{i,i+1}$ describes  couplings between adjacent qubits, and  $\varepsilon_i$ represents the $i$-th on-site energies, with variations caused e.g. by local magnetic fields. Our investigations in this paper focus exclusively on the single excitation subspace of such systems, where we define the state $|1\rangle$ as the excited state.
For a chain of $N$ sites, this single excitation subspace is $N$-dimensional. PST would then be achieved by the state $| \psi_1 \rangle=|10...00\rangle$ being transformed by the system natural dynamics  into the state $| \psi_N \rangle =|00...01\rangle$ after a certain time.

Clearly the evolution of the state is determined by the eigenvalues ($\lambda_{k}$, for $k = 0,1 \dots N-1$) of $H_{XY}$. The success of the pre-selected values of $J_{i,i+1}$ and $\varepsilon_i$, in terms of allowing perfect or quasi-perfect state transfer (PST/QPST), is quantified by the `transfer fidelity' over time
\begin{equation}
F(t) = |\langle \psi_{fin} | e^{-iH_{XY}t/\hbar} | \psi(0) \rangle|^2, \quad 0 \leq F(t) \leq 1
\label{F}
\end{equation}
where $|\psi_{fin}\rangle$ is defined as the desired final state that would achieve PST. Another figure of merit for tracking the success of state transfer protocols, particularly relevant in the field of quantum communications, is the `average fidelity' \cite{Bose_2003, Nikolopoulos2}. This metric is derived by averaging over all pure input states on the Bloch sphere and is given by
\begin{equation}
\langle F_{av}(t) \rangle = \frac{|a_{1,N}(t)|\cos(\nu)}{3} + \frac{|a_{1,N}(t)|^2}{6} + \frac{1}{2},
\label{AvF}
\end{equation}
where \( a_{1,N}(t) = \langle \psi_{\text{fin}} | e^{-iH_{XY}t/\hbar} | \psi(0) \rangle \) is the `transition amplitude' and \( \nu = \text{arg}(a_{1,N}) \). To maximize \( \langle F_{av}(t)\rangle\) across different protocols, a system-specific global magnetic field \( B \) can be applied across the chain, shifting each on-site energy to (\( \epsilon_i + B \)), to ensure that \( \cos(\nu) = 1 \) in Equation~(\ref{AvF}) \cite{Bose_2003}. We note that the fidelity of Equation~(\ref{F}) is effectively the minimum of the set of fidelities that are averaged over to produce this maximised case of Equation~(\ref{AvF}). Thus the fidelity of Equation~(\ref{F}) provides the worst case scenario for QPST with any such given system.

With $| \psi(0) \rangle = | \psi_1 \rangle$, for an $N$-site chain, PST requires $| \psi_{fin} \rangle = | \psi_N \rangle$, the `mirror image' of $|\psi_1\rangle$ with respect to the mirror operator, $M_{ij} = \delta_{i,N+1-j}$, or, in $N\times N$ matrix form
\begin{equation}
    M = \begin{pmatrix}
  0 & 0 & \cdots & 0 & 1 \\
  0 & 0 & \cdots & 1 & 0 \\
  \vdots & \vdots & \ddots & \vdots & \vdots \\
  0 & 1 & \cdots & 0 & 0 \\
  1 & 0 & \cdots & 0 & 0 \\
\end{pmatrix}_{N \times N}
\label{M}.
\end{equation}
For example, 
\begin{equation}
\Hat{M}|\psi_1\rangle = \Hat{M}|1_10_20_3..0_N\rangle = |0_10_20_3..1_N\rangle = |\psi_N\rangle.
\label{m}
\end{equation}
A first step towards ensuring PST dynamics is to choose a mirror-symmetric $H_{XY}$. With $[H_{XY},\Hat{M}]=0$, the energy eigenstates possess even or odd symmetry under $\Hat{M}$. The second step towards ensuring PST is to constrain the spectrum of the mirror-symmetric $H_{XY}$. If the spectrum is such that at a later time $t_m$ (the mirror time) all the odd symmetry eigenstates acquire an additional minus sign in comparison to all the even symmetry eigenstates, the effective evolution operator will be of the form
 $ e^{-iH_{XY}t_m/\hbar} = e^{-i \alpha} \Hat{M}$, for some global dynamical phase $\alpha$. Thus the initial state is mirrored and we may show from Equation (\ref{F}) that \begin{align}
F(t_m) &= |\langle \psi_N| e^{-iH_{XY}t_m/\hbar}| \psi_1 \rangle|^2 \notag \\
     &= |\langle \psi_N| e^{-i \alpha} \Hat{M}| \psi_1 \rangle|^2 \notag \\
     &= |\langle \psi_N| \psi_N \rangle|^2 = 1.
     \label{Ftm}
\end{align}
Therefore, PST is achieved across any $N$-site chain with a mirror-symmetric Hamiltonian and corresponding eigenvalues that fulfil the required condition as outlined above.

Indeed, it is well known \cite{Kay} that the eigenvalues of a system which is capable of complete mirror inversion (PST) Equation (\ref{Ftm}) are restricted to a condition in which the differences between any two sequential eigenvalues $\Delta\lambda_{k,k+1}$, are proportional to a pair-dependent odd integer ($Q_{k,k+1}$) such that
\begin{equation}
    \Delta\lambda_{k,k+1} = \frac{\pi}{t_m} Q_{k,k+1}.
    \label{eig}
\end{equation}
This ensures that, at $t_m$, the dynamical phase factors are what they need to be to effect mirroring of {\it any} initial state.
These considerations are equivalent to previously-established proofs that the ratio of the differences between eigenvalue pairs must be rational \cite{Christandl2, Godsil}. Therefore we may use all this to guide any spectral prescriptions \cite{Kay,Vinet} for PST spin chains.

A further observation, is that our Hamiltonian (Equation (\ref{Hami})) in the case of both uniform coupling and on-site energies, will have eigenvalues that are constrained to a dispersion relation reminiscent of the tight-binding model \cite{Kittel}  
\begin{equation}
    \lambda_k = E - 2J \cos\left(\frac{(k+1)\pi}{N+1}\right), 
    \label{free}
\end{equation}
with $k \in [0,1...,N-1]$ and $E$ a $k$-independent energy. This dispersion relation is very helpful to inform choice of spectra for PST \cite{Yung,Kay,Vinet}. For example, it is clear that Equation (\ref{free}) does not satisfy the conditions of Equation (\ref{eig}). Therefore, if we seek systems with uniform coupling to exhibit PST, we must incorporate non-uniformity into the (mirror-symmetric) profile of on-site energies.

\section{Non-uniform on-site energies ($\varepsilon_i$) in High fidelity QPST solutions}\label{parabolic}
The first approach that we utilise is based on a GA. Here we can impose uniform coupling, and explore optimisation of QPST, as a function of the (mirror-symmetric) onsite energy distribution $\varepsilon_i$.

\subsection{Genetic Algorithm}

Finding solutions which achieve high transfer fidelity across an $N$-site chain can be effectively achieved through evolutionary computation \cite{ES,LC}, with a genetic algorithm as a particularly powerful approach \cite{Luke,BN}. The algorithm used within the following results, begins by generating an initial pool of potential solutions, each assessed through a fitness function that ranks solutions based on their effectiveness. A fitness function, acting as a guide, enables the algorithm to iteratively select the most promising solutions for high-fidelity state transfer and spectral spacing to elicit QPST or PST \cite{BN}. Once a population of prospective solutions is initialized the maximal transfer fidelity score for each individual within the search window is recorded, as well as the relative spacing between its eigenvalues. The highest scoring individuals then `breed' via a crossover function. The offspring are then mutated, with mirror symmetry preserved throughout the mutation process, to subsequently advance to the next generation of prospective candidates.

To maintain a balance between thorough exploration of the solution space and refinement around local optima, mutations alter each generation. These mutations, help prevent premature convergence to suboptimal solutions by broadening the search or fine-tuning specific areas as needed. In this way, the genetic algorithm methodically converges towards solutions that yield the highest fidelity transfer, with the desired spectrum, across the chain.

To identify high-fidelity configurations emerging from a desired spectrum, tuned by a selected $p$-value (discussed within the following subsection), the fitness function is defined as
\begin{equation}
f(F_{\max}, \upsilon(Q, p, \sigma_\lambda)) = \aleph \left[ A \cdot F_{\max} - B \cdot \upsilon(Q, p, \sigma_\lambda) \right],
\label{ff}
\end{equation}
where $F_{\max}$ is the maximum fidelity score achieved within the evaluation period, and $\upsilon$ is the cumulative spectral penalty given by
\begin{equation}
\upsilon = \left| Q - \left( \frac{1}{p} \right) \right| + \sigma_{\lambda},
\label{ups}
\end{equation}
with the $Q$-factor defined as
\begin{equation*}
Q = \frac{\Delta \lambda_{N-2,N-1}}{\left( \prod_{k=0}^{N-3} \Delta \lambda_{k,k+1} \right)^{\frac{1}{N-2}}},
\end{equation*}
and the standard deviation $\sigma_{\lambda}$ of eigenvalue spacings, excluding the highest two levels, given by
\begin{align*}
\sigma_{\lambda} = \sqrt{\frac{1}{N-2} \sum_{k=0}^{N-2} \left(\Delta \lambda_{k,k+1} - \langle \Delta \bar{\lambda} \rangle \right)^2}.
\end{align*}
In Equation (\ref{ff}), $A$ and $B$ are scaling factors that balance fidelity and spectral penalties, while $\aleph$ normalizes the function
\begin{equation*}
\aleph = {(A \cdot F_{\text{max}} + B \cdot \upsilon)}^{-1},
\end{equation*}
ensuring fitness values remain within $[0,1]$. If $A \gg B$, the algorithm prioritizes high-fidelity configurations, whereas for $A \ll B$, spectral characteristics dominate the search.
In each crossover, both parents contribute equally, passing 50\% of their encoding. The offspring then undergo mutation based on a generation-dependent function
\begin{equation}
\mu(g) = \mu_i - g \cdot \frac{\mu_i - \mu_f}{G},
\label{mug}
\end{equation}
where $\mu_i$ and $\mu_f$ are the initial and final mutation rates, respectively. This function is scaled by the total number of generations $G$ and the current generation $g$, ranging from 0 to $G$.
\begin{table}[h]
    \centering
    \renewcommand{\arraystretch}{1.5} 
    \setlength{\tabcolsep}{10pt} 
    \begin{tabular}{|c|c|}
        \hline
         \textbf{Parameter} & \textbf{Value} \\
        \hline\hline
        Number of Generations & 200 \\
        \hline
        Population Size & 1024 \\
        \hline
        Initial Mutation Rate & 20\% \\
        \hline
        Time Window & $t \cdot J_{\max} = 50$ \\
        \hline
    \end{tabular}
    \caption{Genetic Algorithm Parameters, consistent with \cite{BN}.}
    \label{tab:GA_params}
\end{table}The examples discussed in the following section were obtained through genetic algorithm optimization using parameters identical to those originally presented in \cite{BN}, as detailed in Table \ref{tab:GA_params}. 
\subsection{Specific Examples}
\begin{figure}
    \caption*{(a) $N = 4; p = 3$.}
    \makebox[0pt][l]{ 
        \begin{subfigure}[t]{0.45\linewidth}
            \centering
            \includegraphics[width=4.2cm, height=4cm]{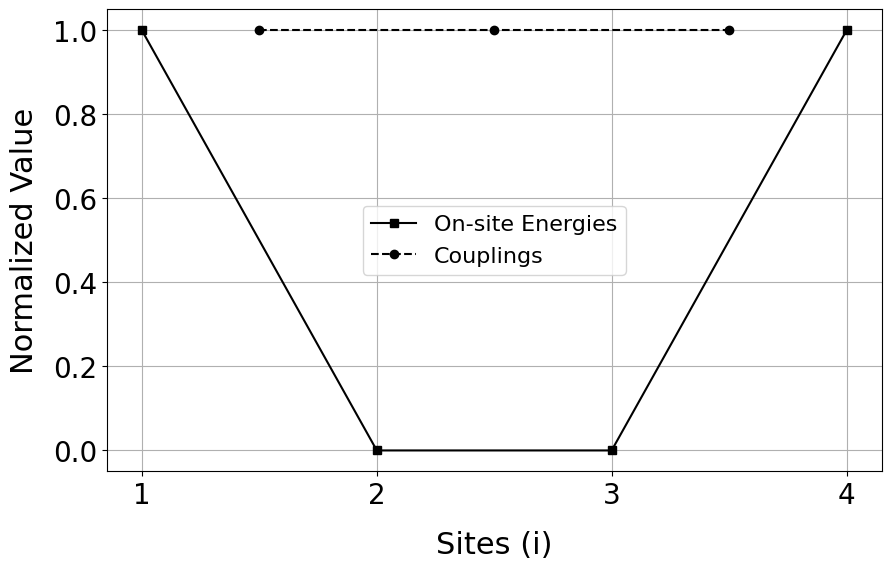}
        \end{subfigure}
    }
    \hspace{-4.6cm} 
    \begin{subfigure}[t]{0.5\linewidth}
        \centering
        \includegraphics[width=4.5cm, height=4cm]{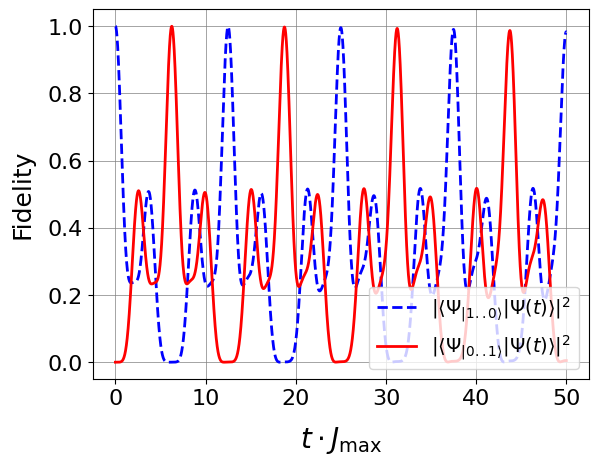}
    \end{subfigure}    
    \vspace{-0.1cm}
    \caption*{(b) $N = 5; p = 3$.}
    \makebox[0pt][l]{ 
        \begin{subfigure}[t]{0.45\linewidth}
            \centering
            \includegraphics[width=4.2cm, height=4cm]{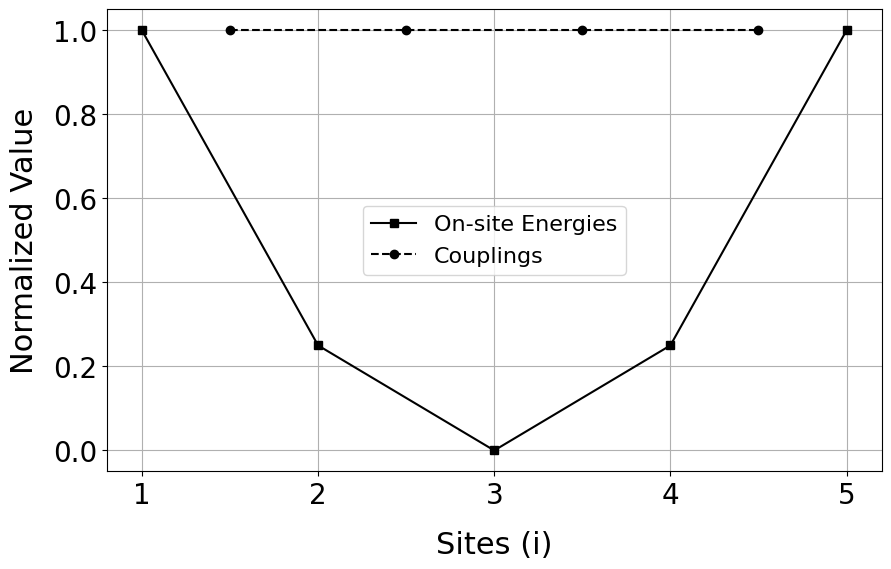}
        \end{subfigure}
    }
    \hspace{-4.6cm} 
    \begin{subfigure}[t]{0.5\linewidth}
        \centering
        \includegraphics[width=4.5cm, height=4cm]{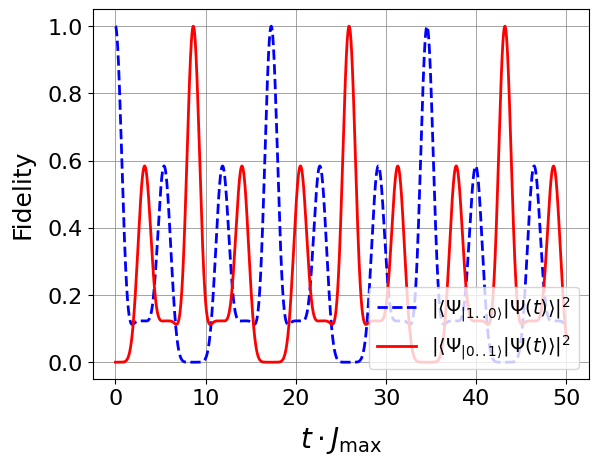}
    \end{subfigure}
    
    \vspace{-0.1cm}
    \caption*{(c) $N = 7; p = 5$.}
       \makebox[0pt][l]{ 
        \begin{subfigure}[t]{0.45\linewidth}
            \centering
            \includegraphics[width=4.2cm, height=4cm]{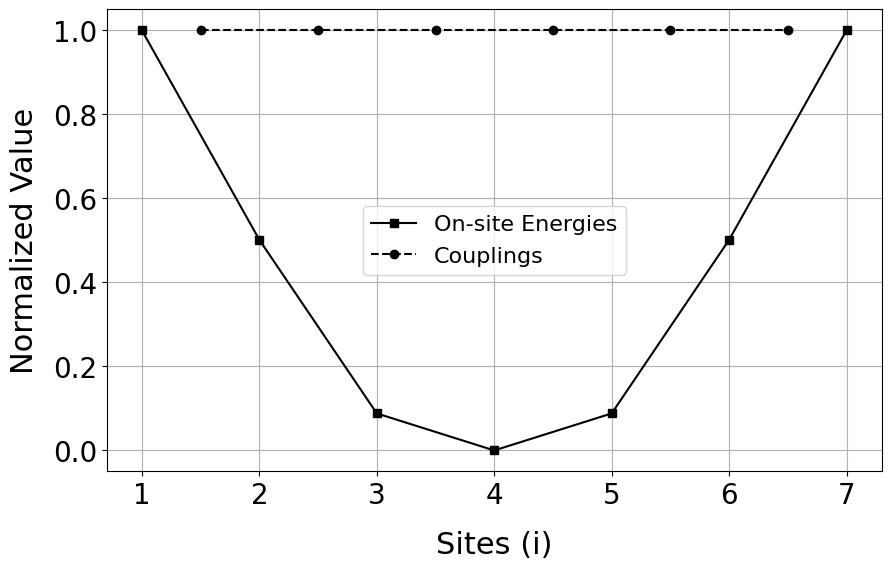}
        \end{subfigure}
    }
    \hspace{-4.6cm} 
    \begin{subfigure}[t]{0.5\linewidth}
        \centering
        \includegraphics[width=4.5cm, height=4cm]{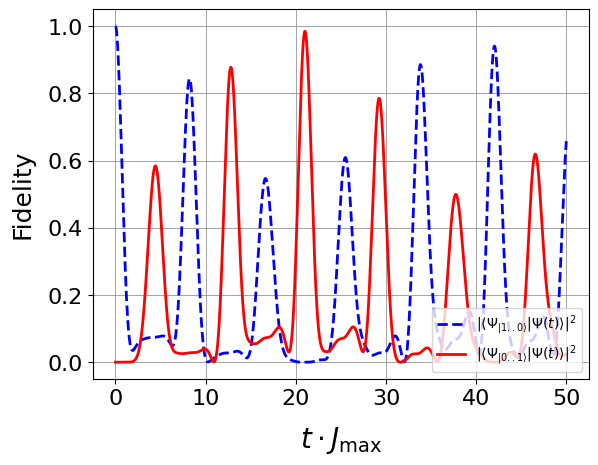}
    \end{subfigure}
        \caption*{(d) $N = 9; p = 9$.}
       \makebox[0pt][l]{ 
        \begin{subfigure}[t]{0.45\linewidth}
            \centering
            \includegraphics[width=4.2cm, height=4cm]{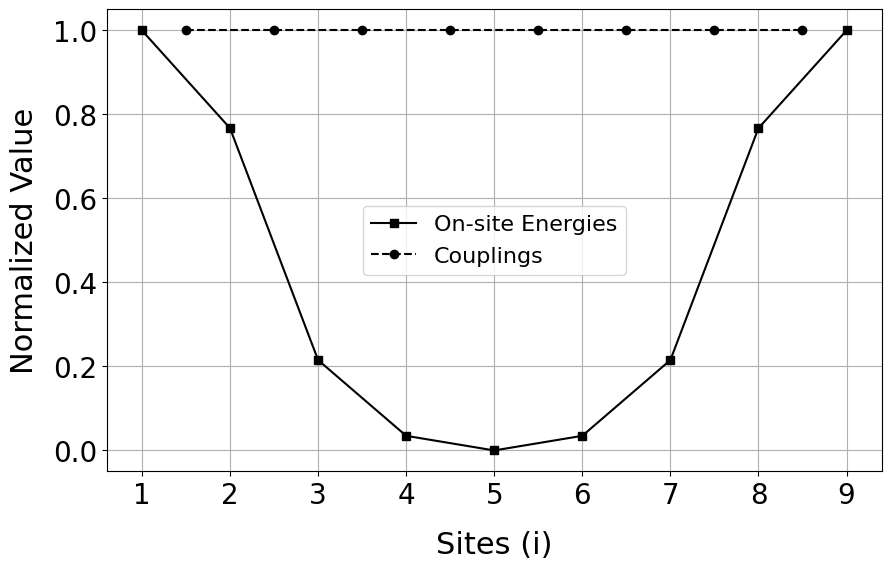}
        \end{subfigure}
    }
    \hspace{-4.6cm} 
    \begin{subfigure}[t]{0.5\linewidth}
        \centering
        \includegraphics[width=4.5cm, height=4cm]{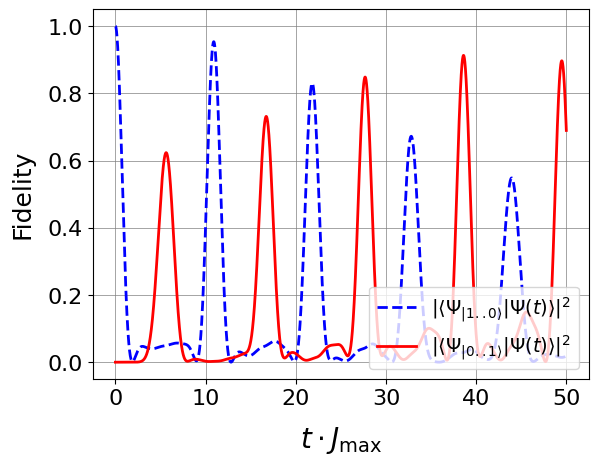}
    \end{subfigure}
\caption{
High-fidelity (transfer fidelity Equation (\ref{F})) dynamics plots (left) for \(N = 4, 5 , 7,\) and \(9\). The corresponding structures of on-site energies and couplings (right) show parabolic energy profiles for homogeneously-coupled chains (solid line with square markers) and mid-bond hopping values for reference (dotted line with circular markers). Y-axis values (right) are normalized as \(\varepsilon_i / \varepsilon_{\max}\).
}

    \label{fig:high-fidelity}
\end{figure}

The results presented here are specific, purposefully chosen examples that demonstrate the decrease in maximal transfer fidelity as chain size increases while maintaining homogeneous couplings, as presented in \cite{BN}.
Fig.  \ref{fig:high-fidelity} illustrates that the observed near-perfect state transfer arises, in part, from the parabolic profile of the on-site energies $\varepsilon_i$. These solutions are found by optimizing for high fidelity transfer given a mostly equidistant then `pinched' spectrum of reciprocal odd integer value,
\begin{equation}
 \lambda_k = \left\{
\begin{array}{ll}
\alpha((1-N) + 2(k-1)), & \text{for}\quad k = 1, 2, \ldots, N-1, \\
\lambda_k = \lambda_{N-1} + \frac{2\alpha}{p}, & \text{for } \quad k = N,
\end{array} \right.
\end{equation}codified by a $p$-factor, between the highest two eigenvalues. The maximal transfer and average fidelity values for these pinched spectra solutions, along with the time achieved, are presented in Table~\ref{tab:transfer_fidelity}. The time values are expressed in dimensionless units of \( t \cdot J_{\max} \), where \( t \) is the evolution time and \( J_{\max} \) is the maximum coupling strength, with \( \hbar = 1 \).
\begin{table}[h!]
    \renewcommand{\arraystretch}{1.7} 
    \setlength{\tabcolsep}{2.5pt} 
    \begin{tabular}{|c|c|c|c|c|}
        \hline
        \textit{\textbf{N}} & Pinch (\(\frac{1}{p}\)) & Max(\(F(t)\)) & Max(\(\langle F_{av}(t) \rangle\)) & Time (\(t \cdot J_{\max}\)) \\ \hline
        4 & \(1/3\) & 0.9997 & 0.9999 & 6.25 \\ \hline
        5 & \(1/3\) & 0.9998 & 0.9999 & 8.63 \\ \hline
        7 & \(1/5\) & 0.9848 & 0.9945 & 21.02 \\ \hline
        9 & \(1/9\) & 0.9128 & 0.9706 & 38.61 \\ \hline
    \end{tabular}
    \caption{Transfer fidelity values for pinched spectra solutions presented in Fig. \ref{fig:high-fidelity}. The pinch ratio \( p \) corresponds to the adjusted spectral compression. Maximum transfer fidelity \( F(t) \) and average fidelity \( \langle F_{av}(t) \rangle \) are calculated over time.}
    \label{tab:transfer_fidelity}
\end{table}

Subsequent quasi-mirroring peaks exhibit modest fidelity decay.  From Fig.   \ref{fig:high-fidelity}(d), we may also see that for a larger system of $N = 9$-sites, with a pinched spectrum of \(\frac{1}{9}\) ($p = 9$), the transfer fidelity has notably reduced and the corresponding structure of the on-site energies (right hand side) have adopted more of an anharmonic configuration to give fidelity scores of $\approx 90\% $. This effect for larger systems sizes $N>10$ is discussed in more detail within Section \ref{sec:IEP} .

Focusing on one of the high-fidelity solutions in the 5-site chain (Fig.\ref{fig:high-fidelity}(a)), which exhibits a mostly equidistant then pinched spectrum, we consider the Hamiltonian
\begin{equation}
H_{XY} = \begin{pmatrix}
3.40 & -0.91 & 0.00 & 0.00 & 0.00 \\
-0.91 & 2.60 & -0.91 & 0.00 & 0.00 \\
0.00 & -0.91 & 2.33 & -0.91 & 0.00 \\
0.00 & 0.00 & -0.91 & 2.60 & -0.91 \\
0.00 & 0.00 & 0.00 & -0.91 & 3.40\\
\end{pmatrix}.
\label{N = 5p3}
\end{equation}
We have identified all the matrix elements $\langle i|H_{ij}|j\rangle$ of Equation(\ref{Hami}), where $|i\rangle$ and $|j\rangle$ denote the excitation's presence on the $i^{\text{th}}$ or $j^{\text{th}}$ site of the chain, respectively. Within Equation (\ref{N = 5p3}), we have simply reversed the sign of all coupling terms relative to their original (positive) values for the purpose of the following discussion. Mathematically, the overall sign of the coupling terms does not affect the system’s dynamics, as long as there are no relative sign differences among them. Two Hamiltonians that differ only in the sign of their off-diagonal coupling terms will exhibit identical temporal evolution, since they share the same eigenvalues \cite{Xie, Chakrabarti}. This adjustment allows for a more direct comparison with alternative formulations explored in subsequent sections. From Fig. (\ref{fig:high-fidelity} (b)), we can see  that though this is a completely homogeneously coupled system, a fidelity of $99.98\%$ can be reached within the time window presented (Table. \ref{tab:transfer_fidelity}). 

\section{Particle in a Discrete Potential analogue}\label{sec:ata}
As mentioned in the introduction, in systems with uniform couplings and non-uniform on-site energies, a large angular momentum ($L$) analogue does not provide the same physical intuition as when the onsite energies are uniform, or zero. Instead, for the systems we study, a discrete model of a particle in a potential can provide some insight. This can be seen by writing the time-independent Schr\"odinger equation, for the Hamiltonian Equation (\ref{Hami}) with fixed negative couplings $J_{i,i+1} = -J$, in component form in the site basis and utilising the discrete difference approximation to the second derivative ($Y^{\prime\prime}_i \equiv \frac{1}{l^2}\left(Y_{i+1} + Y_{i-1} -2Y_i\right)$,
\begin{equation}
    -J l^2 \phi^{\prime\prime}_{k,i} + (\varepsilon_{i} - 2J)\phi_{k,i} = \lambda_k \phi_{k,i} \;.
    \label{Schrod}
\end{equation}
Here $\phi_k$ is the $k-$th eigenstate with eigenvalue $\lambda_k$ ($k=0,1\dots N-1)$, $l$ is the effective length between the sites and of course there are corrections at the chain ends ($i=1$ and $i=N$) due to terms in the second derivative not being present from non-existent sites at $i=0$ and $i=N+1$.

\begin{figure}[h!]
    \centering    \includegraphics[width=0.95\linewidth]{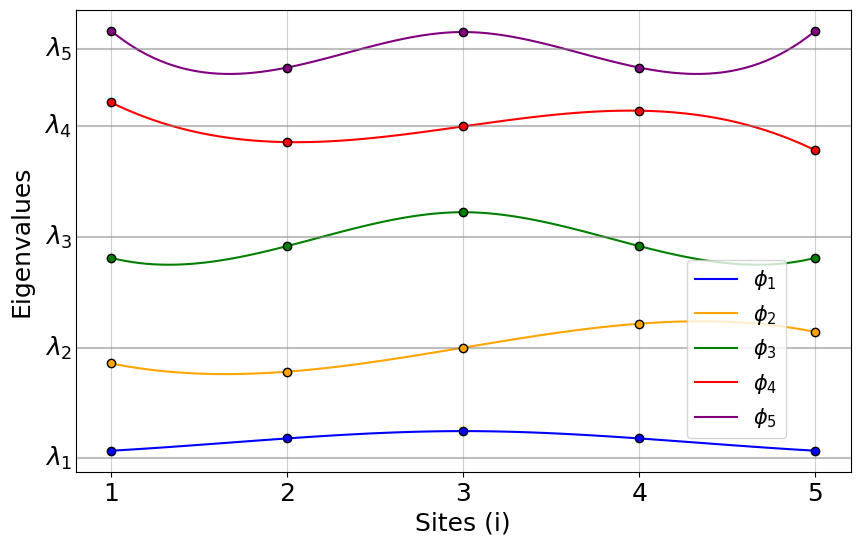}
    \caption{Negative coupling eigenstate amplitude shape, drawn through the actual amplitude points at the discrete sites, (in ascending order, centered about their eigenvalue in the style of an energy-level diagram) of the $N = 5$ high-fidelity solution, plotted against their corresponding eigenvalue, in an energy level diagram-like fashion, along the site indices ($i$).  It can be seen that the eigenstates exhibit forms analogous to those for a harmonic oscillator or other potential well, with $k$ nodes present in the state with energy $\lambda_k$. Correspondingly, the eigenstates alternate in even and odd symmetry (with respect to mirror symmetry about the center of the chain), moving upwards in energy from the even-symmetry ground state. The solid dots serve as a marker for the amplitudes of the eigenstates on site $i$.}
    \label{eigenstates}
\end{figure}

Equation (\ref{Schrod}) shows  an analogue to a particle, with mass inversely proportional to $J l^2$, moving in a discrete potential of $(\varepsilon_{i} - 2J)$. Given the nature of the spectra $\lambda_k$ we seek for PST, this particle analogue gives some physical insight into the forms of the onsite energies $\varepsilon_{i}$ that presented high fidelity QPST in section \ref{parabolic}. These energy profiles are approximately parabolic discrete potential wells, symmetric about the middle of the chain. Such potentials thus provide physical insight into the resultant equally spaced spectra $\lambda_k$, modified only by a pinch of separation between the two uppermost eigenvalues. The physical insight further extends to the forms of the eigenstates, such as in the example given in Fig. (\ref{eigenstates}). The ground state has no node, and the $k-$th excited eigenstate has $k$ nodes, as would be expected for a particle in a potential well.

As can be seen from the largest $N$ example presented in section \ref{parabolic}, with increasing $N$ it starts to become more difficult to maintain both completely uniform chain couplings and high fidelity/QPST. This will be analysed and discussed in detail, with further examples, in the next section. However, prior to that, we continue discussion of the particle in a discrete potential analogy. Even with a relaxation of the coupling uniformity condition, it is still possible to treat the system in terms of raising ($a^{\dagger}$) and lowering ($a$) operators, analogous to those for the simple harmonic oscillator but with some necessary modifications.

We shall focus our attention on systems with spectra known to generate PST, so with an equally spaced spectrum aside from the uppermost separation reduced by a factor of $\frac{1}{p}$ for odd integer $p$, as originally discussed within \cite{BN}. However, the approach is not unique to these spectra, as further modifications to the operators could accommodate other PST spectra. 

If we define the Hamiltonian $\Hat{H}^{\prime}_{XY}$ to be shifted in energy so that its ground state is zero, then we seek operators such that 
\begin{equation}
\Hat{H}^{\prime}_{XY} = \gamma a^{\dagger} a 
\label{Hadaga}
\end{equation}
and with the chosen PST spectrum. This is satisfied with
\begin{eqnarray}
a^{\dagger} = \sqrt{\gamma} \left(\sum^{k=N-3}_{k=0} \sqrt{k+1}\; |\phi_{k+1}\rangle \langle \phi_k |  \right.\nonumber \\  \left. + \;\sqrt{N-2+\frac{1}{p}}\; |\phi_{N-1} \rangle \langle \phi_{N-2}| \right)
\label{adag}
\end{eqnarray}
and correspondingly for $a$, where $\gamma$ is the equal energy spacing of the spectrum and the $|\phi_k\rangle$ (for $k=0,1\dots N-1$) are the energy eigenstate basis. The explicit form of the Hamiltonian
\begin{eqnarray}
\Hat{H}^{\prime}_{XY} = \gamma \left(\sum^{k=N-2}_{k=0} k \;|\phi_{k}\rangle \langle \phi_k |  \right.\nonumber \\  \left. + \;\left(N-2+\frac{1}{p}\right)\; |\phi_{N-1} \rangle \langle \phi_{N-1}| \right)
\label{Hexp}
\end{eqnarray}
clearly demonstrates that the modifications to the operators ensure the correct spectrum. These modifications also guarantee that $a^{\dagger}$, whilst raising from the ground state to produce the other eigenstates, annihilates the uppermost eigenstate $|\phi_{N-1}\rangle$. Furthermore, the commutator 
\begin{eqnarray}
\left[a , a^{\dagger} \right] = \gamma \left( \mathbb{I} - \left( 1 - \frac{1}{p}\right) \;|\phi_{N-2}\rangle \langle \phi_{N-2} |  \right.\nonumber \\  \left. - \;\left(N-1+\frac{1}{p}\right)\; |\phi_{N-1} \rangle \langle \phi_{N-1}| \right)
\label{aadagComm}
\end{eqnarray}
is no longer simply proportional to the identity $\mathbb{I}$, but contains extra terms that enable $a^{\dagger}$ to perform its modified duties as it raises up through the set of eigenstates.

All this demonstrates that it is possible to custom-build a raising and lowering operator picture for these finite systems with spectra that support PST. Another aspect of this picture that can provide further physical intuition is the existence of discrete position $X$ (about the centre of the chain) and momentum $P$ operators. Defining 
\begin{equation}
 a^{\dagger} = X - iP \;\;\;\;a = X +iP \; ,
\label{XPdefs}
\end{equation}
the matrix elements of $X$ in the energy eigenstate basis take the form
\begin{eqnarray}
 \langle \phi_j |X |\phi_k \rangle = \frac{\sqrt{\gamma}}{2} \delta_{j,k+1} \sqrt{k+1 -\left(1-\frac{1}{p}\right)\delta_{k,N-2}}\; \nonumber \\  +\; \frac{\sqrt{\gamma}}{2}\delta_{j+1,k} \sqrt{j+1 -\left(1-\frac{1}{p}\right)\delta_{j,N-2}}\;.
\label{Xmatrix}
\end{eqnarray}
As expected, $X$ connects nearest neighbours in the energy eigenstate basis. In this basis, the mirror operator is diagonal
\begin{equation}
\langle \phi_j |M |\phi_k \rangle = (-1)^{j} \delta_{j,k} \; ,
\label{Menbasis}
\end{equation}
so it is straightforward to see that $X$ and $M$ anticommute, $\{X,M\} = XM + MX = 0$. Given also that $M$ squares to the identity operator $M^2 = \mathbb{I}$, the pairing theorem \cite{PairingTh} applies. Therefore we know immediately that for all $N$ values there exist pairs of $X$ eigenstates, $|x_{+}\rangle$ and $|x_{-}\rangle$, related by $|x_{-}\rangle = M|x_{+}\rangle$ and with eigenvalues related by $x_{-} = - x_{+}$. Furthermore, for odd $N$ values, there exists a zero eigenstate $|x_0\rangle$, which is also an eigenstate of $M$ and has eigenvalue $x_{0} = 0$. This discrete position (from the centre of the chain) representation provides a physical picture for the system, alternative to that provided by the site basis of $|\psi_i\rangle$. Detailed application of this discrete position representation will be presented in future work. Here, we now continue the study of PST and QPST in scenarios where exact coupling uniformity is not enforced.

\section{PST vs. QPST: Persymmetric Matrix Reconstruction}\label{sec:IEP}
We will now provide a method for linear, mirror-symmetric chain construction using inverse eigenvalue methods \cite{Yung, Kay, Vinet, DeBoor, Hochstadt, Hald, Gladwell}: this will allow upgrading of QPST to PST chains. Following this, we also outline some of the potential advantages of the QPST solutions, in comparison to a well-known PST model \cite{Christandl1, Nikolopoulos1} with regard to experimental considerations . 

\subsection{Background}
Since the late 70's, it's been a matter of interest among mathematicians to establish numerical methods/algorithms to reconstruct Jacobi matrices\footnote{Matrices that are positive, semi-definite and tridiagonal with negative co-diagonal elements} from their corresponding spectra. Particularly a Jacobi matrix $H$, of a specific kind, known as `persymmetric' (or mirror-symmetric, $[H_{XY},M] = 0$) offers advantages due to the simplification in computation resulting from the condition that $
\varepsilon_{i} = \varepsilon_{N-i}$ and $
J_{i,i+1} = J_{N - i, N-1+i}$. This is the essence of an inverse eigenvalue problem in which there are a number of established approaches which manipulate monic-orthogonal polynomials \cite{Parlett, Gladwell}, the weights characterized by their discrete inner product, and three-term recurrence relations to yield a numerically stable, and unique persymmetric matrix \cite{DeBoor,Hochstadt,Hald, Gladwell}. Examples of this method, applied to spin chains, have been presented by Yung and Bose \cite{Yung}, and by Kay and Vinet \cite{Kay, Vinet}. The novelty of our approach is to apply the inverse eigenvalue method of persymmetric matrix reconstruction to systematically extrapolate trends for the parameters of the resultant Hamiltonians obtained from spectra which satisfy PST, and doing this over progressively larger systems (See Fig. \ref{Standarddeviation}). Furthermore, we use the inverse eigenvalue method to `round' individual eigenvalues found from optimization techniques to the nearest numerical values, ensuring that Equation(\ref{eig}) is satisfied. This process highlights structural differences between QPST solutions and their adjacent PST counterparts.

\begin{figure}
    \begin{minipage}{.85\linewidth}
        \centering
        \includegraphics[width=1.0\linewidth]{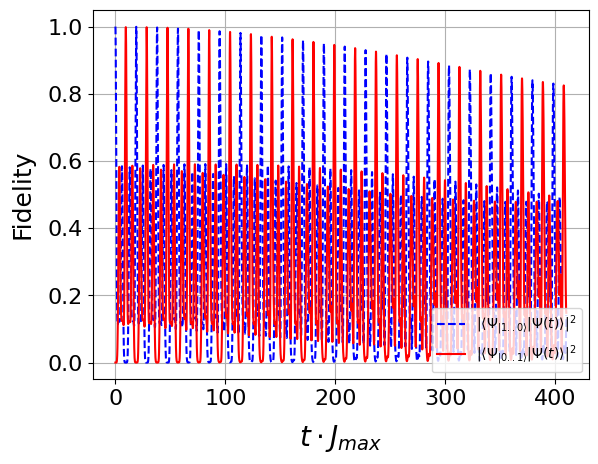} 
    \end{minipage}
    
    \vspace{-1.cm} 
    
    \begin{minipage}{0.85\linewidth}
        \centering
        \includegraphics[width=1.0\linewidth]{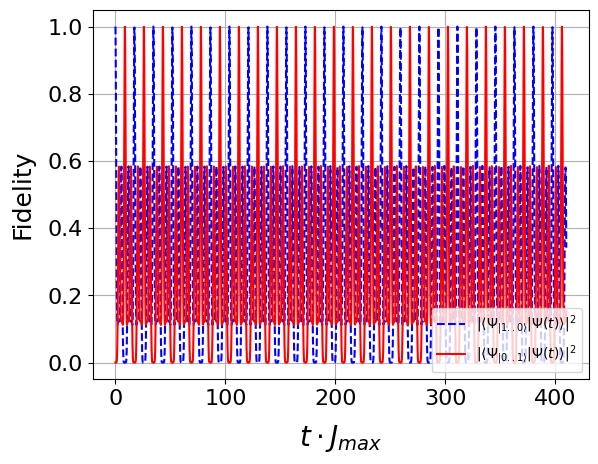} 
    \end{minipage}
    \caption{High-fidelity dynamics of the 5-site chain described by Equation \ref{N = 5p3} (top plot) plotted over an extended window. The transfer fidelity tends towards (approximately) $80\%$ within a window of $t \cdot J_{max} = 400$. Transfer fidelity over extended time window for $H^*_{XY}$ (lower plot) reconstructed using the inverse eigenvalue method. It can be seen through plotting over larger time scales that the transfer fidelity associated with the correctly chosen spectral data for the persymmetric matrix reconstruction remains arbitrarily close to 1 and periodic.}
    \label{N5ext}
\end{figure}

As discussed previously, the spectral condition of Equation~(\ref{eig}) must be fulfilled for perfect state transfer to occur. Furthermore, it has also been shown that if the eigenvalues satisfy the condition 
\begin{equation}
\lambda_k = -\lambda_{N-1-k}
\label{symm}
\end{equation}
for all $k$, such that the spectrum is symmetric about its center, then all $\varepsilon_i$ are equal to a constant $C$, which can 
be effectively translated to all on-site energies being zero \cite{Kay}.
The high transfer fidelity observed in Fig.~\ref{fig:high-fidelity}(b) is achieved through a non-symmetric spectrum, thereby violating Equation~(\ref{symm}). However, this fidelity deteriorates over time, becoming noticeably weaker when the observation window of the state's evolution is significantly extended (see Fig.~\ref{N5ext}).

\subsection{Generating unique $\varepsilon_i$ and $J_{i,i+1}$ values from a chosen spectrum $\{\lambda_k\}$}
Let $P_N(\lambda) = \prod_{k=1}^{N} (\lambda - \lambda_k)$ represent the characteristic polynomial of $H$, with the $j$th leading principal minor of the matrix $(\lambda I - H_N$), denoted by $P_j(\lambda)$, representing the characteristic polynomial of truncated (starting from the upper left corner) $H$. The sequence of polynomials $P_j(\lambda)$ forms a Sturm sequence, satisfying a three-term recurrence relation for $j = 1, 2, \ldots, N$
\begin{equation}
P_j(\lambda) = (\lambda - \epsilon_j) P_{j-1}(\lambda) - J_{j-1}^2 P_{j-2}(\lambda).
\label{recur}
\end{equation}
A Sturm sequence, by definition, exhibits a crucial property, in that the roots of $P_j(\lambda)$, denoted as $\lambda_{jk}$, interlace those of $P_{j-1}(\lambda)$
\begin{equation}
\lambda_{j,j-1} < \lambda_{j-1,j-2} < \lambda_{j,j-2} < \cdots < \lambda_{j,1} < \lambda_{j-1,0} < \lambda_{j,0}.
\label{interlace}
\end{equation}
This property leads to the following relationship \cite{Yung}
\begin{equation*}
\text{sgn}[P_{N-1}(\lambda_k)] = (-1) \times \text{sgn}[P_{N-1}(\lambda_{k-1})],
\end{equation*}
which serves as mathematical foundation of the alternating parity of the eigenvectors, as these polynomials are related to the coefficients within the eigenbasis expansion. Using a combined approach, only requiring knowledge of the spacing of the spectral data to calculate the weights \cite{DeBoor, Hochstadt}
\begin{equation}
w_k = \prod_{\substack{j \neq k}}^{N-1} \frac{1}{\lvert \lambda_k - \lambda_j \rvert}
\label{w}
\end{equation}
which are defined with respect to orthogonal polynomials discrete inner products via
\[
\langle f, g \rangle = \sum_{k=1}^{N} f(\lambda_k) g(\lambda_k) w_k,
\]
where $f(\lambda_k)$ and $g(\lambda_k)$ are generic monic-orthogonal polynomials, we may capitalize on the three-term recurrence relation and interlacing property of the roots of the polynomials (Equation(\ref{interlace})) to deduce the entries of the tridiagonal matrix.\footnote{Note, two polynomials are said to be orthogonal with respect to the weights $w_k$, which measures the contribution of specific eigenvalues to their inner product \cite{Gladwell}.}
We may use the recurrence relation (Equation(\ref{recur})) and initially set $P_0$ to $1$, and $P_1(\lambda_k)$ to $\lambda_k - \varepsilon_1$, where $\lambda_1$ is one of the eigenvalues of $H_N$, to find the on-site potentials
\\
\begin{align}
\epsilon_{i} &=  \frac{\langle \lambda \cdot P_{i-1,:}, P_{i-1,:} \rangle}{\lVert P_{i-1,:} \rVert^2} =  \frac{\sum_{k=1}^{N} \lambda_k \cdot P_{i-1}(\lambda_k)^2 \cdot w_k}{\sum_{k=1}^{N} P_{i-1}(\lambda_k)^2 \cdot w_k},
\label{var}
\end{align}
and the couplings
\begin{align}
J_{i,i+1} &= \frac{\lVert P_{i,:} \rVert}{\lVert P_{i-1,:} \rVert} = \frac{\sum_{k=1}^{N} P_{i}(\lambda_k)^2 \cdot w_k}{\sum_{k=1}^{N} P_{i-1}(\lambda_k)^2 \cdot w_k},
\label{coupling}
\end{align}
governed by $\lambda_k \in \{ \lambda_1, \lambda_2, \ldots, \lambda_N \}$ \cite{Gladwell}. For further details on the inverse eigenvalue method, the reader is referred to Appendix~\ref{appen}, which includes a worked example for $N = 3$.
The polynomials evaluated at each eigenvalue $P_i(\lambda_k)$, which satisfy Equation  (\ref{recur}), and the weights $w_k$, maintains accurate representation of the spectral structure within the stable numerical reconstruction \cite{DeBoor}.
\subsection{Transitioning from QPST to PST}
Through direct diagonalisation of the matrix Equation \ref{N = 5p3}, the associated spectrum which dictates the success of transfer throughout the time evolution of the state, can be shown to nearly but not perfectly satisfy the conditions of Equation \ref{eig}, i.e. 
\begin{equation}
\lambda_{N=5} \in \left\{1.006, 2.006, 3.001, 3.994, 4.326\right\}.
\end{equation}
Therefore the peak-values of the transfer fidelity revivals decay over time, see upper panel of Fig. \ref{N5ext}.  We may however use the inverse eigenvalue method to reconstruct the unique persymmetric matrix which fulfills the PST conditions by using the spectrum 
\begin{equation}
\lambda^*_{N=5} \in \left\{1.00, 2.00, 3.00, 4.00, 4.3\overline{3}\right\}.
\label{lambda}
\end{equation}
 The solution we find for the system capable of PST corresponding to Equation (\ref{lambda}) is then
\begin{equation}
\small
H^*_{XY} = \begin{pmatrix}
3.40 & -0.9165 & 0.00 & 0.00 & 0.00 \\
-0.9165 & 2.60 & -0.9129 & 0.00 & 0.00 \\
0.00 & -0.9129 & 2.33 & -0.9129 & 0.00 \\
0.00 & 0.00 & -0.9129 & 2.60 & -0.9165 \\
0.00 & 0.00 & 0.00 & -0.9165 & 3.40
\end{pmatrix}.
\label{N = 5rec}
\end{equation}
 While this spectral reconstruction analysis centers on the specific example of a 5-site chain, the QPST configurations presented both here and throughout \cite{BN} can undergo similar treatment, allowing for the recovery of their respective PST matrix elements.

\begin{figure}[h!]
    \centering
    \begin{minipage}{0.75\linewidth}
        \centering
        \includegraphics[width=\linewidth]{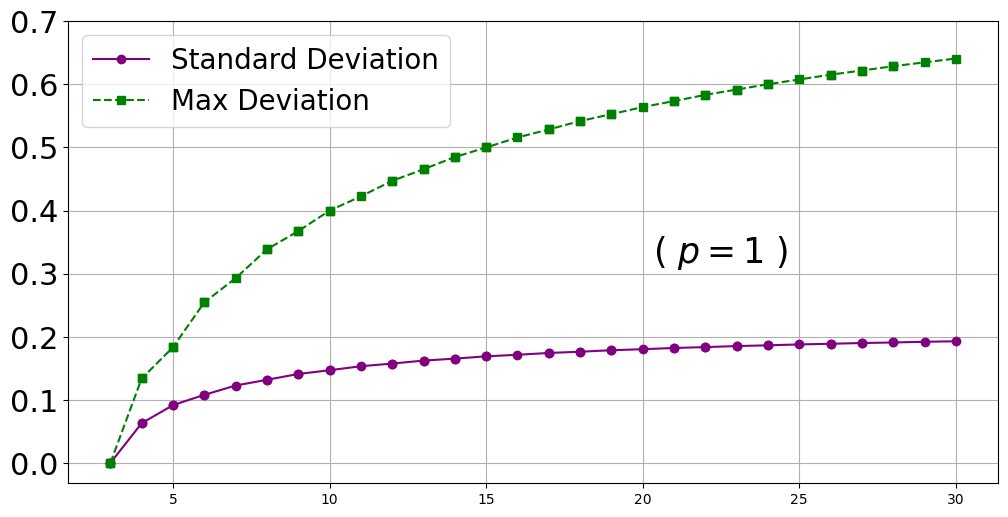}
    \end{minipage}
    
    \vspace{    -0.25cm} 
    
    \begin{minipage}{0.75\linewidth}
        \centering
          
        \includegraphics[width=\linewidth]{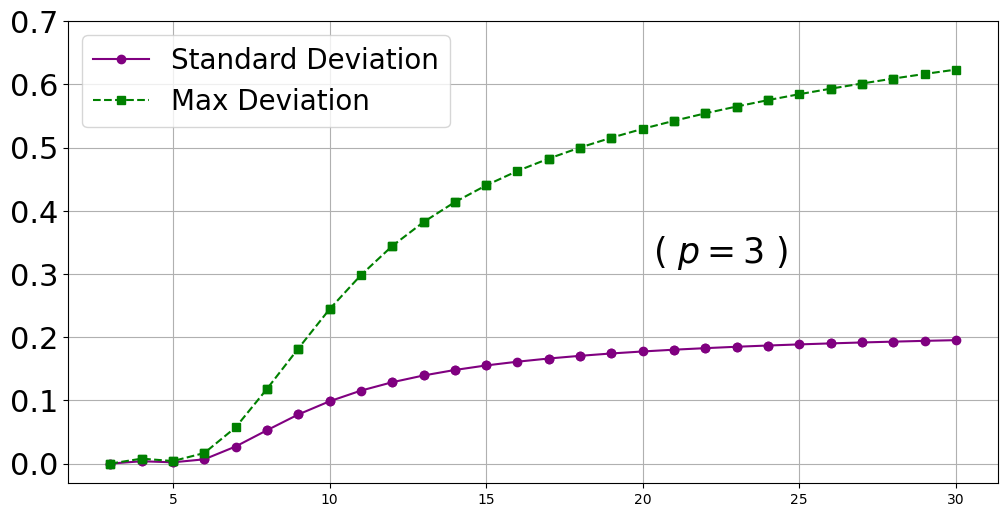}
    \end{minipage}
    
    \vspace{    -0.25cm} 
    
    \begin{minipage}{0.75\linewidth}
        \centering
        \includegraphics[width=\linewidth]{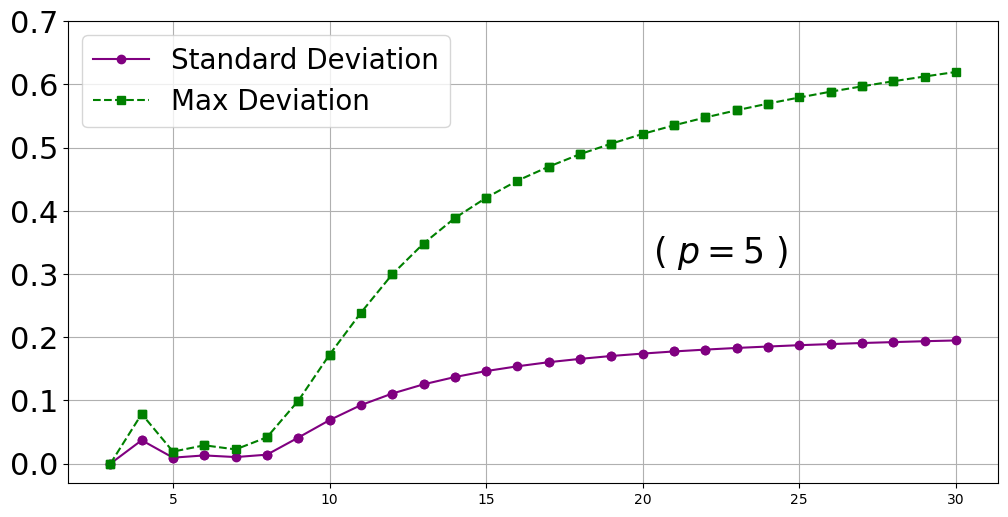}
    \end{minipage}
    
    \vspace{    -0.25cm} 
    
    \begin{minipage}{0.75\linewidth}
        \centering
          
        \includegraphics[width=\linewidth]{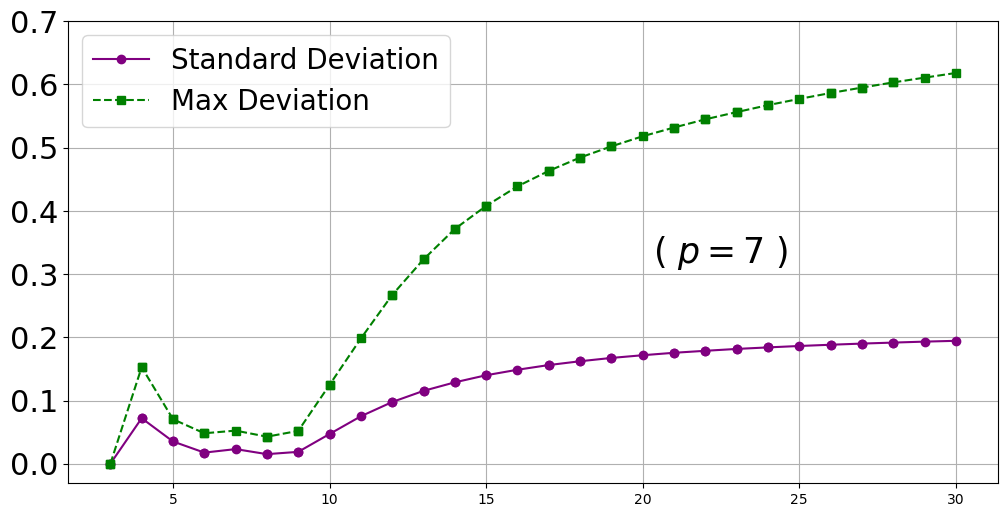}
    \end{minipage}
    
    \vspace{    -0.25cm} 
    
    \begin{minipage}{0.75\linewidth}
        \centering
         
        \includegraphics[width=\linewidth]{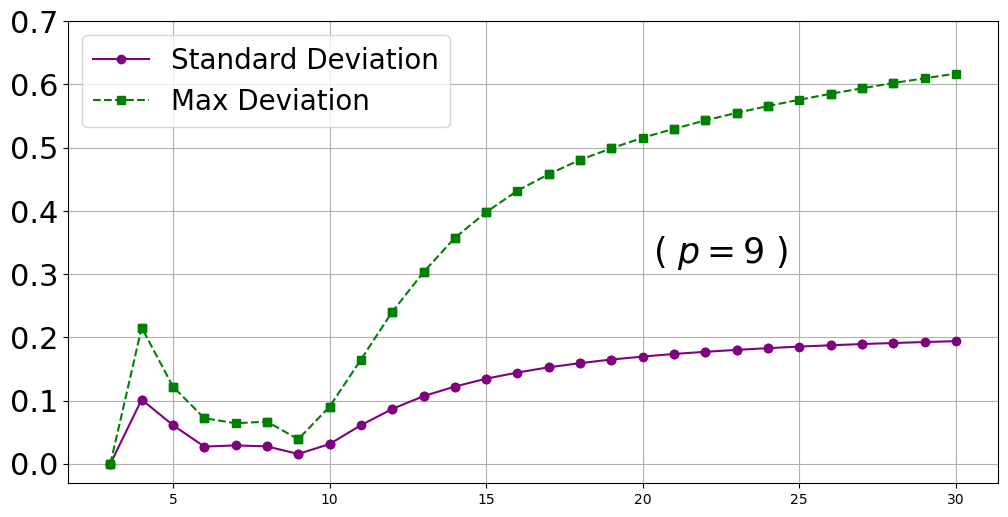}
    \end{minipage}
    
    \vspace{    -0.25cm} 
    
    \begin{minipage}{0.75\linewidth}
        \centering
          
        \includegraphics[width=\linewidth]{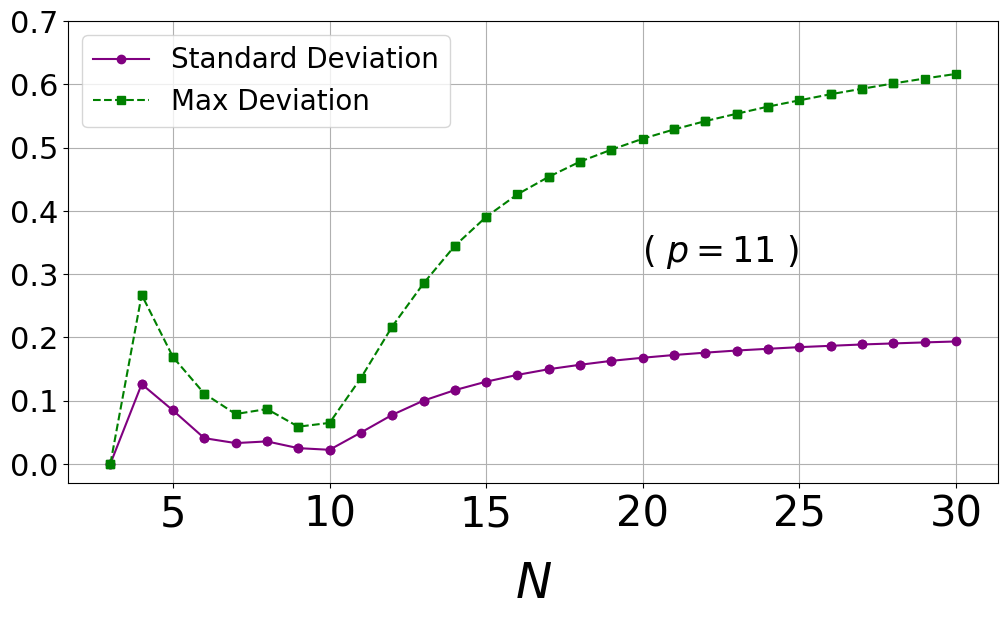}
    \end{minipage}

     \caption{The standard and maximum deviation numerical values (y-axis) in the couplings as a function of $N$, derived from inverse spectral persymmetric matrix construction.}
    \label{Standarddeviation}
\end{figure}

Though the true PST configurations do not harbor complete homogeneity amongst the coupling terms (Equation (\ref{N = 5rec})), and therefore capitalizing on the apparent experimental advantages of such solutions, the difference in the energies of the coupling required to transition from QPST to PST is just 0.4\%. In comparison, the well-known  coupling model \cite{Christandl1, Nikolopoulos1}
\begin{equation}
   J_{i,i+1} = J_0\sqrt{(N-i)i}, 
   \label{christa}
\end{equation}
with initial strength $J_0 = 1$, yields a difference of $\approx 20\%$, for $N = 5$, between the maximum and minimum coupling strengths throughout the chain for PST to occur. The standard deviation in the coupling, as a function of $N$, due to this coupling model (Equation (\ref{christa})) may be observed within Fig.   \ref{Standarddeviation}. 

We also note that, for this non-equidistant pinched spectrum reconstructed via the persymmetric matrix method, both the standard deviation of the coupling values and the maximum deviation between the largest and smallest couplings exhibit a distinct trend. As shown in Fig.~\ref{Standarddeviation}, this deviation forms a `well' for $N\lesssim10$, as $p$ increases, but tends to saturate for larger $N$-systems. This provides an explanation as to why QPST solutions could be found for smaller $N$-site chains using a tailored optimization algorithm, but as the system sizes grow that it becomes progressively more difficult to find high-fidelity solutions which correspond to practically/nearly homogeneously-coupled chains.

\section{Experimental Implementations}
Several qubit hardware platforms enable precise control over both on-site energies and coupling values, allowing for fine-tuned manipulation of individual sites and their interactions. Semiconductor and superconducting qubits, employ electrostatic and magnetic gates to externally control on-site energy tuning and qubit-qubit coupling \cite{Xiang, Botzem}. In ion trap arrays, the on-site energies may be effectively modulated through axial vibrational mode cooling and non-uniform magnetic field gradients, which induce site-specific Zeeman shifts, enabling resonance frequency tuning at the individual ionic level \cite{W2,W3}. Similarly, cold atom arrays utilize site-dependent external potentials to modify on-site energies, as a well-established technique in these systems \cite{Bugnion2013, Le2023}.

This extensive and varied range of realisations for spin chains provides a firm experimental foundation for our proposed approach. In some of the platforms discussed, inter-qubit couplings may not be as easily tunable and are instead approximated as constant. For such hardware, we have demonstrated that QPST is attainable. However, if more precise control over couplings is available, PST can also be achieved by adjusting the couplings as illustrated in Fig.~\ref{Standarddeviation}. For a more detailed discussion on experimental implementations and perturbative effects, we refer the reader to \cite{BN}.
\section{Conclusions}
To conclude, this work highlights the feasibility and potential advantages of achieving high-fidelity state transfer in linear spin chains with non-uniform on-site energies. By incorporating the different methodologies of genetic algorithms and inverse eigenvalue approaches, we demonstrated the ability to optimize spin chain configurations for quasi-perfect and perfect state transfer, respectively. The analogy to a quantum particle in a discrete potential provides valuable physical insights, particularly in the context of uniform coupling and tailored on-site energy profiles. Whilst quasi-perfect state transfer solutions offer practical advantages for smaller chain sizes, the transition to perfect state transfer via persymmetric matrix reconstruction becomes increasingly critical for larger systems to maintain high fidelity. These findings underscore the interplay between spectral constraints, coupling uniformity, and on-site energy distributions, contributing to the broader goal of efficient quantum information processing. Future work will pursue further refinement of these approaches, particularly for larger systems or additional experimental considerations.
\[\]
\noindent C.C. Nelmes acknowledges support from EPSRC, grant number is EP/W524657/1.
\section*{Appendix} \label{sec:appendix}
\subsection{The Persymmetric Matrix Spectral Reconstruction Method: An $N=3$ Example}
\label{appen}
To explicitly demonstrate the iterative procedure, we consider the spectrum
\[
\lambda_k \in \{1, 2,2+\frac{1}{p}\}
\]
which is consistent with the `pinch' spectra originally discussed within \cite{BN}. As we understand the spectrum, we may therefore compute the weights using Equation (\ref{w})
\begin{align*}
w_1 &= \frac{p}{p + 1},\quad w_2 = p,  \quad w_3 =\frac{p^2}{p + 1}.
\end{align*}
We know from the previous subsection that $P_0(\lambda_k) = 1$
and therefore using Equation (\ref{var}) we now compute the first diagonal entry
\begin{align*}
\varepsilon_1 &= \frac{\sum_{k=1}^3 P_0(\lambda_k)^2\cdot\lambda_k \cdot w_k}{\sum_{k=1}^3 w_k} \\\\
&= \frac{(\lambda_1)\cdot w_1 + (\lambda_2)\cdot w_2 + (\lambda_3)\cdot w_3}{\frac{p}{p + 1} + p + \frac{p^2}{p + 1}}\\\\
&=\frac{(1) \cdot \frac{p}{p + 1} + (2)\cdot p + \left(2 + \frac{1}{p}\right)\cdot \frac{p^2}{p + 1}}{2p}
= 2.
\end{align*}
Through obtaining the value of $\varepsilon_1$ we may now evaluate the first polynomial $P_1(\lambda_k) = \lambda_k - 2$ for each eigenvalue 
\begin{align*}
P_1(\lambda_1) &= 1 - 2 = -1 \\
P_1(\lambda_2) &= 2 - 2 = 0 \\
P_1(\lambda_3) &= \left(2 + \frac{1}{p}\right) - 2 = \frac{1}{p}
\end{align*}
and subsequently compute the first coupling term (from Equation (\ref{coupling})) to be
\begin{align*}
J_{1,2}^2 &= \frac{\sum_{k=1}^3 P_1(\lambda_k)^2 \cdot w_k}{\sum_{k=1}^3w_k}\\
&= \frac{(P_1(\lambda_1))^2 \cdot w_1 + (P_1(\lambda_3))^2 \cdot w_3}{2p}\\
&= \frac{(-1)^2 \cdot \frac{p}{p + 1} + (\frac{1}{p})^2 \cdot \frac{p^2}{p + 1}}{2p}\\
&=  \frac{1}{2p}.
\end{align*}
As we now have the values of $P_1(\lambda_k)$ for each eigenvalue and also found (from the previous step) $\sum_{k=1}^3 P_1(\lambda_k)^2 \cdot w_k =1$ we now compute the second main diagonal entry of the persymmetric matrix to be
\begin{align*}
\hspace{1cm} \varepsilon_2 &= \frac{\sum_{k=1}^3 \lambda_k \cdot P_1(\lambda_k)^2 \cdot w_k}{\sum_{k=1}^3 P_1(\lambda_k)^2 \cdot w_k} \\
&= (1) \cdot (-1)^2\cdot \frac{p}{p + 1} + \left(2 + \frac{1}{p}\right) \cdot (\frac{1}{p})^2 \cdot \frac{p^2}{p + 1}\\
&= \frac{p + 2 + \frac{1}{p}}{p + 1}.
\end{align*}
One may optionally proceed to compute \( P_2(\lambda_k) \) using the recurrence relation in order to obtain the final diagonal entry \( \varepsilon_3 \). However, for the persymmetric case with \( N = 3 \), we know that
\[
\varepsilon_1 = \varepsilon_3 \quad \text{and} \quad J_{1,2} = J_{2,3},
\]
so the Jacobi matrix is now fully determined. The final (unique) output matrix is therefore
\begin{equation*}
H_{XY} = \begin{pmatrix}
2 & \frac{1}{\sqrt{2p}} & 0 \\
\frac{1}{\sqrt{2p}} & \frac{p + 2 + \frac{1}{p}}{p + 1} & \frac{1}{\sqrt{2p}}\\
0 & \frac{1}{\sqrt{2p}} & 2 \\ 
\end{pmatrix}
\end{equation*}
which can be diagonalized and verified (for chosen $p-$values) as having the initial spectrum of $\lambda_k \in \{1,2,2+\frac{1}{p}
\}$. 

Necessary for larger $N$-site chains $(N\geq5)$, the next step in the iteration is to use the recurrence relation (Equation (\ref{recur})) to derive the expression for the next sequential polynomial 
\[
P_2(\lambda_k) = (\lambda_k-\varepsilon_2)P_1(\lambda_k)-J_{1,2}^2P_0,
\]evaluate it for each eigenvalue, and derive the next main diagonal $\varepsilon_3$ and off-diagonal $J_{2,3}$ terms via its computation.

\end{document}